\documentclass[aps,prl,twocolumn,groupedaddress,reprint]{revtex4-1}
\usepackage{graphicx}
\usepackage{setspace}
\usepackage{amsmath}
\usepackage{amssymb}
\usepackage{color}
\usepackage{array}
\usepackage{subfigure}
\usepackage{hyperref}
\usepackage{float}
\usepackage{lipsum}

\begin{document}
%\begin{CJK*}{GB}{}
\title{Precisely resolve energy-time entanglement by dual channel Fabry-P\'{e}rot interferometry}
\author{Yuan Sun}
\affiliation{Department of Physics and Astronomy, Stony Brook University, Stony Brook, NY 11794--3800}

\begin{abstract}
Precisely resolving the continuous variable energy-time entanglement of paired photons is an important issue in quantum optics. The Fabry-P\'{e}rot interferometer provides a distinguished opportunity for this purpose if the single photon pulse's self-interference is carefully avoided. A dual channel Fabry-P\'{e}rot interferometer is proposed and studied with the focus put upon higher order quantum interference effects. When the two channels are properly set up, it is capable of resolving the energy-time entanglement in detail analogously to that a usual Fabry-P\'{e}rot interferometer can resolve classical light's spectrum. Variation form of the dual channel Fabry-P\'{e}rot interferometry is also discussed.
\end{abstract}
\maketitle
%\end{CJK*}

Energy-time entanglement, being an outstanding example of the entanglement between continuous variables \cite{nphys2492, PhysRevLett.84.5304}, has aroused the enthusiasm of many researchers in recent years. It provides convincing evidence against the local hidden variable (LHV) theory \cite{PhysRevLett.110.260407}, and has potential important applications, such as the quantum cryptography and the quantum key distribution \cite{PhysRevLett.84.4737, PhysRevLett.98.060503, PhysRevLett.112.120506}. Ever since the Franson interferometry \cite{PhysRevLett.62.2205}, many interesting schemes \cite{PhysRevLett.101.180405, PhysRevLett.102.040401, 1367-2630-16-1-013033} have been proposed to examine the energy-time entanglement experimentally. Researchers have been working hard to improve the fringe visibilities \cite{PhysRevA.47.R2472, PhysRevLett.66.1142, PhysRevLett.93.010503, nature07121} for better contrast in the violation of the Bell inequality. Also, the highly non-local properties of energy-time entanglement have been investigated by various interferometric methods \cite{PhysRevLett.65.321, PhysRevLett.81.3563, PhysRevA.87.053822}.
\par
Intuitively, if two photons are energy-time entangled, the collapse of one photon's wave packet onto some eigenstate specified by the detection methods at some time will necessarily mandate the time at which the other photon's wave packet collapses \cite{nphys2492}. Very often the total energy of such a photon pair is conserved with an uncertainty much less than that of an individual party. Usually, the energy-time entangled biphoton posses an experimental signature of strong correlations in the time at which they are registered as ``clicks'' in the detectors and their frequencies \cite{PhysRevLett.113.063602}. However, this type of correlations alone is not enough to serve as the proof for energy-time entanglement, nor does it provide the full information of the entanglement. On the other hand, two-photon Fabry-P\'{e}rot interferometer for time-bin entanglement \cite{Stucki2005a} has already been studied thoroughly and we wonder whether an analogous interferometry exists for the very different entanglement of energy and time.
\par
%make this paragraph strong.
In this letter, we propose a dual channel Fabry-P\'{e}rot interferometer that has the ability to precisely resolve the energy-time entanglement of photon pairs. The aim of the proposed interferometry is not only to determine the existence of energy-time entanglement, but also to allow the actual experimental measurement of the fine structure of the energy-time entanglement in an unambiguous way. Since many of the potential quantum optics applications of the energy-time entanglement rely upon its essentially unlimited Hilbert space volume pertinent to the continuous variable entanglement, a method to precisely map out the entanglement is certainly of help. 
\par
\begin{figure}[t!]
 \centering
\begin{tabular}{l}
\includegraphics[trim = 0mm 20mm 0mm 65mm, clip, width=9cm]{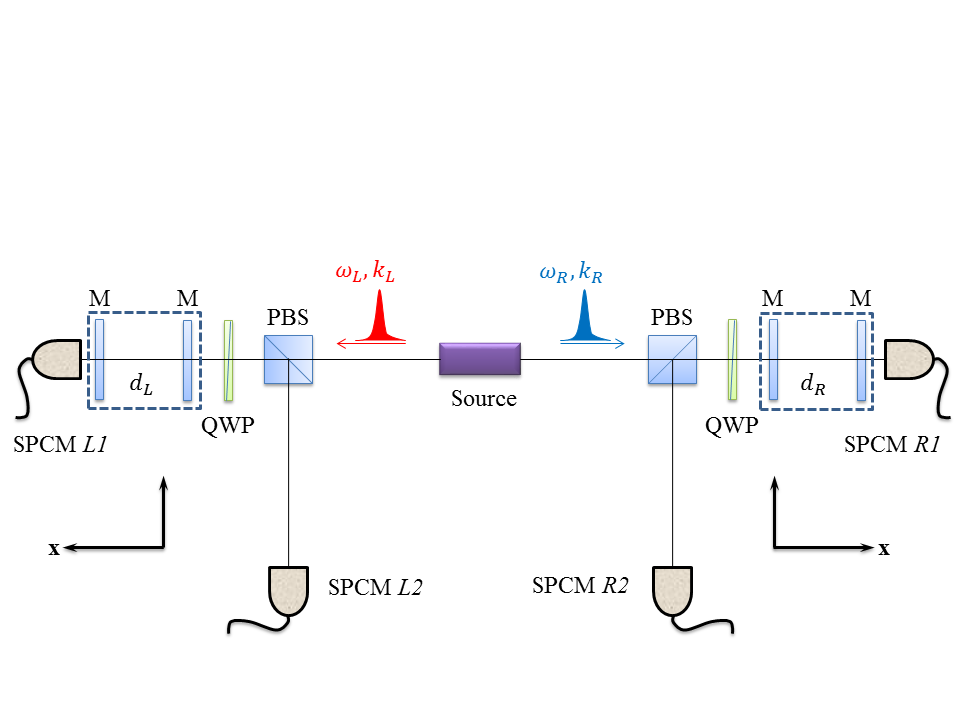}
%left down right up
\end{tabular}
\linespread{1} 
\caption{(Color online) Schematic of the equal arm dual channel Fabry-P\'{e}rot interferometer. Abbreviations in the graph: mirror (M); quarter--wave plate (QWP); polarization beam splitter (PBS). The source emits energy--time entangled photon pair traveling along the two symmetric arms separately, where the polarizations are assumed to be fixed. The positive direction of each arm's local coordinate system is along the $k$--vector of the photons propagating in that arm. The polarizations of the emitted photons are such that they make it through when they first hit the PBS. In both arms the transmissions and the reflections are monitored by SPCM. The parts in the dashed box are individual Fabry-P\'{e}rot interferometers. They have similar cavity lengths $d_L \approx d_R$ in the sense that $Q\times|d_L-d_R|$ is small compared to the individual photon pulse length ($c\times \text{single photon coherence time}$) where $Q$ is the quality factor of the Fabry-P\'{e}rot interferometers. Here the Fabry-P\'{e}rot interferometers are composed of planar mirrors for the ease of derivations, while in real experiments concave mirrors shall be considered to make them work in the stable resonator regime. If only the experimental results of one arm are collected, it would look purely normal as a single photon pulse experiment and no knowledge about the other arm can be gained.\label{schematic_1}}
\end{figure}

The setup of this interferometer is sketched in Fig. \ref{schematic_1} . It is composed of two arms of equal length, where a Fabry-P\'{e}rot interferometer is inserted into each arm. The two Fabry-P\'{e}rot interferometers have similar cavity lengths and all their mirrors are the same: planar, with field transmission coefficient $T$ and field reflection coefficient $R$  satisfying $T^2 + R^2 =1$. Each arm is equipped with two single photon counting modules (SPCM) with assumed 100\% quantum efficiency. A single photon pulse $\Psi(kx-\omega t) |0\rangle$ is to be registered as a click at time $t$ satisfying $kx-\omega t = 0$ if the SPCM is placed at position $x$, where $\Psi$ is the field operator generating the single photon pulse and $|0\rangle$ is the vacuum state. A key requirement is that the individual single photon pulse's coherence time is less than the round trip time $2d_{L,R}/c$ in the cavities in order to avoid self-interferences. The biphoton source is postulated to emit energy-time entangled photon pairs: a subscript $L$ or $R$ is attached to distinguish the photons coming down the left or right arm.
%The entanglement is regarded as perfect, immune to decoherence effects and lasting forever.
\par
The discussion is carried out under the Heisenberg picture, where the field operator evolves in time.  Suppose that when a pair of energy-time entangled photons is generated in free space, the field operator is given as in Eq.\eqref{initial_state}. Eq.\eqref{initial_state} describes a photon pair that is perfectly energy-time entangled while both photons have only a single frequency component. All the counts registered by the SPCM $L1$ \& $R1$ are coincidence counts in this case. Eq.\eqref{initial_state} is regarded as an idealization to make the derivations simpler \cite{SuppInfo}.
\begin{align}
\label{initial_state}
\Phi_\text{total} = \int  e^{i(k_Lx-\omega_L t)}\Psi_L(k_Lx-\omega_L(t-\tau)) \nonumber\\
 \times e^{i(k_Rx-\omega_R t)}\Psi_R(k_Rx-\omega_R(t-\tau)) d\tau.
\end{align}
\par
When the two Fabry-P\'{e}rot interferometers are inserted into the two arms separately as in Fig. \ref{schematic_1}, higher order quantum interference occurs for the two-photon state $\Phi_\text{total} |0\rangle$. First we turn our attention to the transmission that will induce coincidence counts in SPCM $L1$ and $R1$, which is essentially a form of post-selection. The field operator $\Phi^T_\text{coind}$ describing such a transmission state is given in Eq.\eqref{coind_1}.
\begin{align}
\label{coind_1}
&\Phi^T_\text{coind}
= \int T^4 \sum_{l=0}^\infty R^{4l}\nonumber\\
&\times e^{i(k_R(x+2d_Ll) - \omega_L t)}\Psi_L (k_L(x+2d_Ll) - \omega_L (t-\tau))  \nonumber\\
&\times e^{i(k_R(x+2d_Rl) - \omega_R t)}\Psi_R (k_R(x+2d_Rl) - \omega_R (t-\tau)) d\tau.
\end{align}
\par
We make the following observation as in Eq.\eqref{observation_1}, which is the consequence of energy--time entanglement \cite{SuppInfo}.
\begin{align}
\label{observation_1}
&\int \Psi_L (k_L(x+2d_Ll) - \omega_L (t-\tau)) \nonumber\\
&\qquad \times\Psi_R (k_R(x+2d_Rl) - \omega_R (t-\tau)) d\tau \nonumber\\
&= \int \Psi_L (k_L x - \omega_L (t-\tau))\Psi_R( k_R x - \omega_R (t-\tau)) d\tau, \forall l.
\end{align}
\par
Combining Eq.\eqref{coind_1} and Eq.\eqref{observation_1}:
\begin{align}
\label{coind_2}
&\Phi^T_\text{coind}
= \frac{T^4}{1 - R^4 e^{i 2 k_L d_L + i 2 k_R d_R}} e^{i(k_Lx-\omega_L t)} e^{i(k_Rx-\omega_R t)} \nonumber\\
&\times \int \Psi_L (k_L x - \omega_L (t-\tau))\Psi_R( k_R x - \omega_R (t-\tau)) d\tau.
\end{align}
\par
\begin{figure}[h]
 \centering
\begin{tabular}{l}
\includegraphics[trim = 0mm 4mm 0mm 0mm, clip, width=9cm]{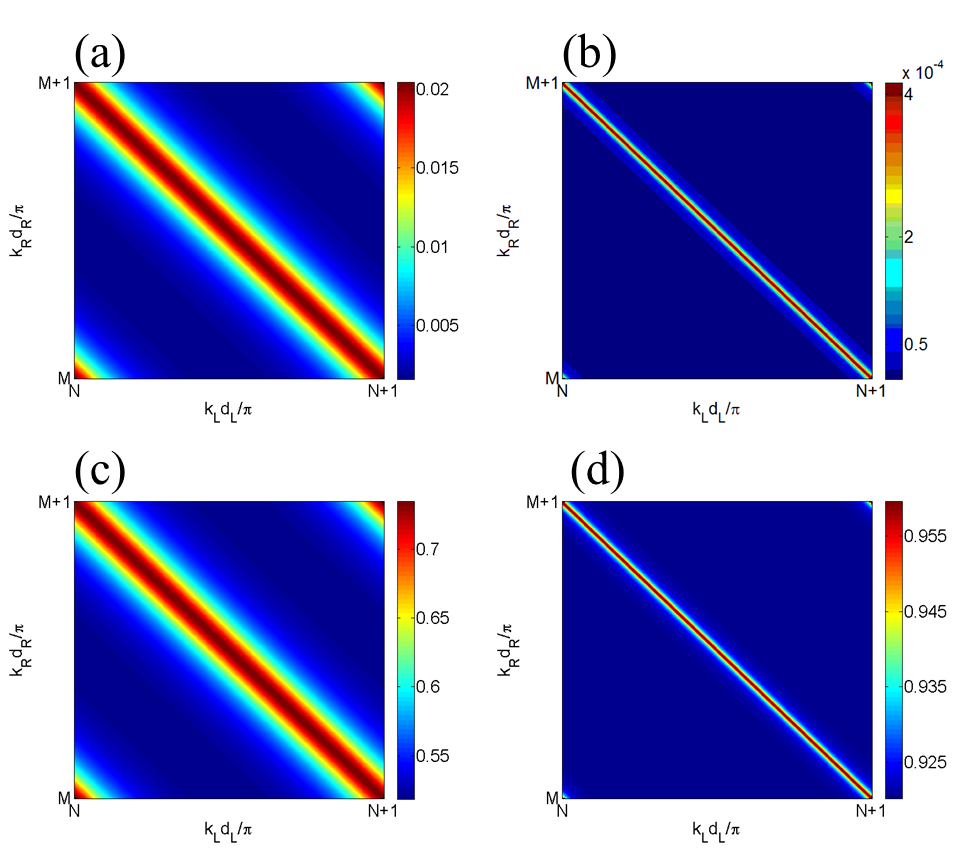}
%left down right up
\end{tabular}
\linespread{1} 
\caption{(Color online) Numerical simulation results of the transmission and reflection coincidence rates for the interferometer in Fig. \ref{schematic_1}. All rates are normalized to the coincidence rate without inserting the two Fabry-P\'{e}rot interferometers. The rates are plotted as a function of the Fabry-P\'{e}rot cavity lengths $d_{L,R}$ where $M, N$ are integer numbers. (a) transmission coincidence for $T = 0.5$; (b) transmission coincidence for $T = 0.2$; (c) reflection coincidence for $T = 0.5$; (d) reflection coincidence for $T = 0.2$; \label{num_siml_1}}
\end{figure}

From Eq.\eqref{coind_2} it is clear that the transmission coincidence rate is mediated by the factor $T^4/(1 - R^4 e^{i 2 k_L d_L + i 2 k_R d_R})$, which is analogous to the usual Fabry-P\'{e}rot transmission and leads to sharp ``resonances''. The transmission coincidence rate can be calculated as in Eq.\eqref{coind_rates_1}, when normalized to the coincidence rate if the two Fabry-P\'{e}rot interferometers are removed:
\begin{equation}
\label{coind_rates_1}
\text{rates} = \frac{T^8}{1+R^8 - 2R^4 \cos(2k_L d_L + 2k_R d_R)} .
\end{equation}
The ``on--resonance'' condition that leads to the maximum transmission coincidence is given in the following Eq.\eqref{on_resonance_1}.
\begin{equation}
\label{on_resonance_1}
k_L d_L + k_R d_R = N\pi,
\end{equation}
where $N$ is some positive integer number. Fig. \ref{num_siml_1} (a) \& (b) present numerical simulation results for the transmission coincidence rates for $T=0.5$ and $T=0.2$.
\par
Eq.\eqref{coind_2} and Eq.\eqref{coind_rates_1} are highly non-local results coming from higher order quantum interference as a consequence of the energy-time entanglement. Suppose that no entanglement exists, even if the pair of photons is always generated at the same time, the transmission coincidence counting rate would not vary with respect to the tiny change in the cavity length (change at the scale $k_{L,R} \Delta d_{L,R} \sim \pi$). Moreover when entanglement does exist, Eq.\eqref{coind_rates_1} serves as the fully quantum mechanical prediction which refutes the prediction of an usual LHV model \cite{SuppInfo, PhysRevA.45.8138, PhysRevA.61.012105}.
\par
In the above discussions the frequencies of the two photons are specified. However it is not necessary to demand the photons to have fixed frequencies in order to make this interferometer behave. An adequate condition is that $\omega_L + \omega_R = \text{conserved}$ under the already mentioned postulation that $d_L \approx d_R$, and then our analysis from Eq.\eqref{coind_1} to Eq.\eqref{on_resonance_1} is still valid. This observation implies that this type of interferometry has the ability to detect the energy conservation of the biphotons within the accuracy of the inserted Fabry-P\'{e}rot interferometer's linewidth, provided the fine experimental control over $d_{L,R}$.
\par
The transmission coincidence makes up only a small part of all the transmissions and such a post--selection process would necessarily throw away a lot of SPCM's counts. More often, the clicks in SPCM $L1$ and $R1$ for the transmission will be separated by some time interval. For simplicity we put $d_L = d_R = d$ for the moment. Then the possible separation time between a click in $L1$ and a click in $R1$ is $\Delta t = m\cdot2d/c, m=0, \pm1, \pm2, \cdots$ where positive $m$ means a click at $L1$ is ahead of a click at $R1$ while negative $m$ means a click at $R1$ is ahead of a click at $L1$. The biphoton field operator $\Phi^T_m$ governing the transmission for a specific interval $\Delta t(m)$ is therefore calculated in Eq.\eqref{transmission_m_1}.
\begin{align}
\label{transmission_m_1}
&\Phi^T_m =  \frac{T^4 R^{2|m|} e^{-i\omega_L \Delta t}}{1 - R^4 e^{i 2 k_L d + i 2 k_R d}} e^{i(k_Lx-\omega_L t)} e^{i(k_Rx-\omega_R t)} \nonumber\\
&\times \int \Psi_L (k_L x - \omega_L (t+\Delta t-\tau))\Psi_R( k_R x - \omega_R (t-\tau)) d\tau.
\end{align}
\par
Eq.\eqref{transmission_m_1} covers all the possibilities of the post--selection process for the transmission counts' correlations. In other words, the total transmission biphoton field operator $\Phi^T_\text{total}$ can be understood as $\Phi^T_\text{total} = \sum_{m=-\infty}^{m=+\infty} \Phi^T_m$. Eq.\eqref{transmission_m_1} implies that the non--coincidence counts are not just wastes but also carry vital information about the quantum interference as a result of the entanglement.
\par
We now extend our analysis to the reflections registered by SPCM $L2$ and $R2$. First the field operator $\Phi^R_\text{coind}$ for reflection coincidence is computed in Eq.\eqref{refl_coind_1}.
\begin{align}
\label{refl_coind_1}
&\Phi^R_\text{coind}
=  \frac{R^2(1 - (R^2-T^2)e^{i 2 k_L d + i 2 k_R d})}{1 - R^4e^{i 2 k_L d_L + i 2 k_R d_R}} \nonumber \\
&\qquad \times e^{i(k_Lx-\omega_L t)} e^{i(k_Rx-\omega_R t)} \nonumber\\
&\times \int \Psi_L (k_L x - \omega_L (t-\tau))\Psi_R( k_R x - \omega_R (t-\tau)) d\tau.
\end{align}
which has an interesting tiny difference compared to the usual form of Fabry-P\'{e}rot's reflection. Namely, the minus sign from the half wave loss disappears because in the two photon case it is multiplied twice. Fig. \ref{num_siml_1} (c) \& (d) present numerical simulation results for the reflection coincidence rates.
\par
Not all the reflection counts on SPCM $L2$ and $R2$ are coincidences either. Similarly, assume $d_L = d_R = d$ for the moment. If the separation time between a click in $L2$ and a click in $R2$ is $\Delta t = m\cdot2d/c, m=\pm1, \pm2, \cdots$, then the corresponding biphoton field operator $\Phi^R_m$ is computed in the following Eq.\eqref{refl_m_1}.
\begin{align}
\label{refl_m_1}
&\Phi^R_m =  \frac{R^2T^2R^{2|m|-2}(-1+R^2e^{i 2 k_L d + i 2 k_R d}) e^{-i\omega_L \Delta t}}{1 - R^4 e^{i 2 k_L d + i 2 k_R d}} \nonumber\\
&\qquad \times e^{i(k_Lx-\omega_L t)} e^{i(k_Rx-\omega_R t)} \nonumber\\
&\times \int \Psi_L (k_L x - \omega_L (t+\Delta t-\tau))\Psi_R( k_R x - \omega_R (t-\tau)) d\tau.
\end{align}
\par
The correlation between the counts of SPCM $L2$ and $R1$ is also subject to the same quantum interferences. Keeping the definition of $\Delta t$ and the assumption of $d_L = d_R = d$ as above, we construct the biphoton field operator $\Psi^{RT}_m$ for the post--selection that a click in $L2$ occurs at $\Delta t(m)$ ahead of $R1$. When $m = 0, 1,2,\cdots$, $\Psi^{RT}_m$ is as in Eq.\eqref{mix_m_1}, while when $m = -1,-2,-3,\cdots$, $\Psi^{RT}_m$ is as in Eq.\eqref{mix_m_2}.
\begin{align}
\label{mix_m_1}
&\Phi^{RT}_m
=  \frac{T^2R^{2m+1}(-1 + R^2e^{i 2 k_L d + i 2 k_R d})e^{-i\omega_L \Delta t}}{1 - R^4e^{i 2 k_L d_L + i 2 k_R d_R}} \nonumber \\
&\qquad \times e^{i(k_Lx-\omega_L t)} e^{i(k_Rx-\omega_R t)} \nonumber\\
&\times \int \Psi_L (k_L x - \omega_L (t+\Delta t-\tau))\Psi_R( k_R x - \omega_R (t-\tau)) d\tau,
\end{align}
\begin{align}
\label{mix_m_2}
&\Phi^{RT}_m =  \frac{T^4 R^{2m-1} e^{-i\omega_L \Delta t}}{1 - R^4 e^{i 2 k_L d + i 2 k_R d}} e^{i(k_Lx-\omega_L t)} e^{i(k_Rx-\omega_R t)} \nonumber\\
&\times \int \Psi_L (k_L x - \omega_L (t+\Delta t-\tau))\Psi_R( k_R x - \omega_R (t-\tau)) d\tau.
\end{align}
\par
\begin{figure}[h]
 \centering
\begin{tabular}{l}
\includegraphics[trim = 55mm 60mm 0mm 15mm, clip, width=9cm]{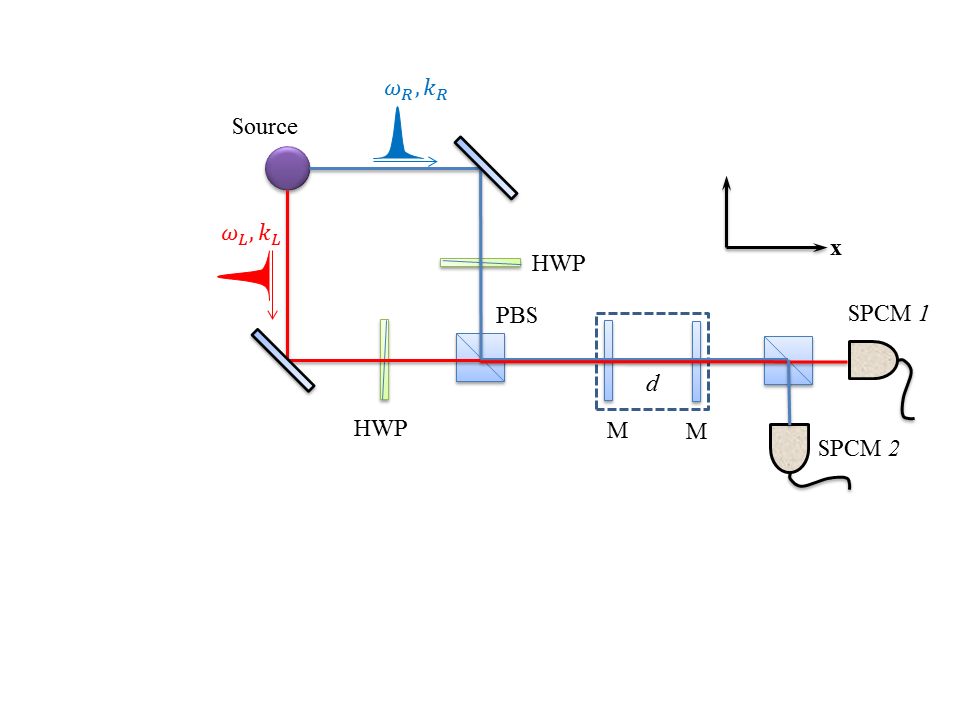}
%left down right up
\end{tabular}
\linespread{1} 
\caption{(Color online) Dual channel interferometry realized via two polarization modes in a single Fabry-P\'{e}rot interferometer. HWP stands for half wave plate. A pair of photons with orthogonal polarizations are combined, sent through the same cavity in the same spacial mode and then split towards the SPCM's after they transmit. It shall become clear that the dual channels can be realized by a lot of methods, such as spatial separation, different polarizations, or even different spatial modes.\label{schematic_2}}
\end{figure}

This type of dual channel interferometry has close ties with weak values \cite{RevModPhys.86.307}, and its description can be switched to an equivalent weak value formulation. It also shares some basic ideas with the geometric phases \cite{PhysRevLett.101.180405, PhysRevLett.91.090405}. It has a variation form whose schematic is presented in Fig. \ref{schematic_2}. By manipulating polarizations, dual channel interferometry for the energy--time entanglement is realized with only one actual Fabry-P\'{e}rot interferometer.
\par
\begin{figure}[h]
 \centering
\begin{tabular}{l}
\includegraphics[trim = 0mm 0mm 0mm 0mm, clip, width=9cm]{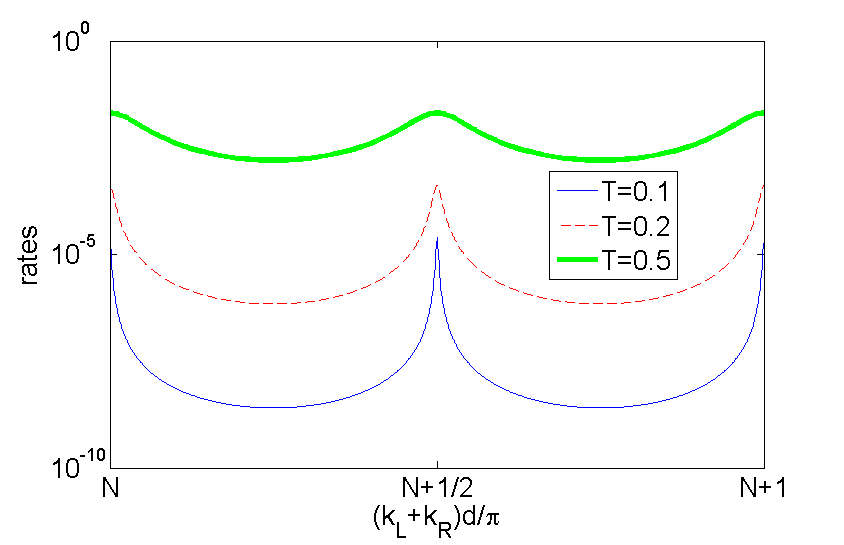}
%left down right up
\end{tabular}
\linespread{1} 
\caption{(Color online) Numerical simulation of the transmission coincidence rates for the interferometer in Fig. \ref{schematic_2} with different values of $T$. The rate is plotted as a function of the cavity length $d$, in logarithmic scale. All rates are normalized to the coincidence rate without inserting the Fabry-P\'{e}rot interferometer. \label{num_siml_2}}
\end{figure}

The design in Fig. \ref{schematic_2} works on the same principle as that in Fig. \ref{schematic_1} and yet has its distinct features. The artificial requirement of $d_L \approx d_R$ or $d_L = d_R = d$ is now automatically fulfilled. Eq. \eqref{transmission_m_1} applies here without modification. A numerical simulation for its performance in terms of coincidence rates is given in Fig. \ref{num_siml_2}, where the signal strength drops significantly as the reflectivity of the cavity mirror increases. One way to gather more signal to counteract this drawback is through the utilization of all the $m$'s of Eq.\ref{transmission_m_1}, and this method is similar to a recent method used to enhance the signal in the N00N state imaging experiment \cite{PhysRevLett.112.223602}.
\par
For the sake of a succinct discussion, the assumed form of the entangled biphoton in Eq.\eqref{initial_state} does not adequately address the intricate details of the energy--time entanglement. A simple but more realistic model is provided in Eq.\eqref{mod_state}, where $\varphi(\tau)$ contains the information about the energy--time entanglement.
\begin{align}
\label{mod_state}
\Phi_\text{total} = \int \varphi(\tau)  e^{i(k_Lx-\omega_L t)}\Psi_L(k_Lx-\omega_L(t-\tau)) \nonumber\\
 \times e^{i(k_Rx-\omega_R t)}\Psi_R(k_Rx-\omega_R(t-\tau)) d\tau.
\end{align}
\par
An immediate observation is that Eq.\eqref{mod_state} leads to energy uncertainties in the photons which Eq.\eqref{initial_state} does not. The transmission coincidence of the biphoton in the form of Eq.\eqref{mod_state} through the dual channel Fabry-P\'{e}rot interferometer is then computed in Eq.\eqref{mod_coind}.
\begin{align}
\label{mod_coind}
&\Phi^T_\text{coind}
= T^4e^{i(k_Lx-\omega_L t)} e^{i(k_Rx-\omega_R t)} \nonumber\\
&\qquad\times\int\sum_{l=0}^{\infty} R^{4l}e^{i2l(k_L+k_R)d} \varphi(\tau+2l\frac{d}{c})  \nonumber\\
&\times \Psi_L (k_L x - \omega_L (t-\tau))\Psi_R( k_R x - \omega_R (t-\tau)) d\tau.
\end{align}
\par
Provided the condition that $\Psi_L$ and $\Psi_R$ represent short single photon pulses, Eq.\eqref{mod_coind} means that the coincidence rate as a function of $d$ is essentially a windowed Fourier transform of $\varphi(\tau)$. Therefore an experimental measurement similar to Fig. \ref{num_siml_1} or Fig. \ref{num_siml_2} will yield the spectral property of $\varphi(\tau)$. Through this model analysis it is clear that the dual channel Fabry-P\'{e}rot interferometry has the ability to analyze the biphoton energy--time entanglement in a very analogous way that a usual Fabry-P\'{e}rot reveals the incident classical light's frequency spectrum.
\par
We discuss briefly three factors that will compromise the performance of this type of interferometry: phase noise, environmental noise and quantum detection efficiency. The phase noise of the optical field introduces unwanted stochasticity just like its role in the laser--atom interaction \cite{PhysRevA.89.032516}. Environmental noise causes decoherence and may even lead to entanglement sudden death \cite{PhysRevLett.97.140403}. A low quantum detection efficiency will impede the data acquisition and smear the interference pattern's visibility.
\par
In conclusion, we have proposed a dual channel Fabry-P\'erot interferometer and discussed in theory its properties. We have shown that it has the ability to precisely analyze the energy-time entanglement between a pair of photons and therefore it has the potential to become a standard tool for such tasks.

\begin{acknowledgments}
The author acknowledges support from ONR and help from Dr. Harold Metcalf, which makes this work possible.
\end{acknowledgments}

\bibliographystyle{apsrev4-1}

\renewcommand{\baselinestretch}{1}
\normalsize

%\clearpage%
%\phantomsection%
%\addcontentsline{toc}{chapter}{\numberline{}{Bibliography}}%
\bibliography{ysref}
\end{document}